\documentclass{elsart}

\usepackage{graphicx}

\usepackage{amssymb}

\newcommand{\dsp}{\displaystyle}
\newcommand{\Fig}[1]{Fig.~\ref{#1}}
\newcommand{\Figs}[1]{Figs.~\ref{#1}}

\newcommand{\Eq}[1]{Eq.~(\ref{#1})}
\newcommand{\Eqs}[2]{Eqs.~(\ref{#1}) and (\ref{#2})}

\newcommand{\mytheta}{\theta}

\newcommand{\Thlabi}[1]{\ensuremath{\mytheta_{\rm lab,#1}}}
\newcommand{\Thlabx}{\ensuremath{\mytheta_{\rm lab}}}

\newcommand{\cms}{c.m.}

\newcommand{\gcm}{\ensuremath{\gamma_{\rm \cms}}}

\newcommand{\ddb}{\ensuremath{\bar{d}}}
\newcommand{\dds}{\ensuremath{\sigma_d}}
\newcommand{\ddp}{\ensuremath{\phi_d}}
\newcommand{\delp}{\ensuremath{\delta\phi}}
\newcommand{\xb}{\ensuremath{x_{b}}}
\newcommand{\sxb}{\ensuremath{\delta x_{b,FWHM}}}

\newcommand{\yb}{\ensuremath{y_{b}}}
\newcommand{\xp}{\ensuremath{x'_{b}}}
\newcommand{\yp}{\ensuremath{y'_{b}}}
\newcommand{\xps}{\ensuremath{x^{'2}_{b}}}
\newcommand{\yps}{\ensuremath{y^{'2}_{b}}}
\newcommand{\xt}{\ensuremath{\tilde{x}}}
\newcommand{\yt}{\ensuremath{\tilde{y}}}
\newcommand{\zt}{\ensuremath{\tilde{z}}}
\newcommand{\mmm}{\ensuremath{\mathcal{M}}}
\newcommand{\mytextwidth}{\textwidth}
\newcommand{\CH}{\ensuremath{{\rm CH}_2}}

\newcommand{\grad}{\ensuremath{^{\circ}}\ }
\newcommand{\gradx}{\ensuremath{^{\circ}}}
\newcommand{\chiz}{\ensuremath{\chi^2}}

\begin{document}

\begin{frontmatter}

\title{Determining Beam Parameters in a Storage Ring 
with a Cylindrical Hodoscope using Elastic Proton-Proton
Scattering \thanksref{support}}

\thanks[support]{Supported by the BMBF and FZ J\"ulich}


\author[ISKP]{ H.~Rohdje\ss\corauthref{cor1}}
\ead{rohdjess@iskp.uni-bonn.de}
\author[HH]{ D.~Albers} 
\author[ISKP]{ J.~Bisplinghoff} 
\author[HH]{ R.~Bollmann} 
\author[HH]{ K.~B\"u\ss{}er} 
\author[ISKP]{ O.~Diehl} 
\author[HH]{ F.~Dohrmann} 
\author[ISKP]{ H.-P.~Engelhardt} 
\author[ISKP]{ P.~D.~Eversheim} 
\author[HH]{ M.~Gasthuber} 
\author[HH]{ J.~Greiff} 
\author[HH]{ A.~Gro\ss{}} 
\author[ISKP]{ R.~Gro\ss{}-Hardt} 
\author[ISKP]{ F.~Hinterberger}
\author[HH]{ M.~Igelbrink} 
\author[HH]{ R.~Langkau} 
\author[IKP]{ R.~Maier}
\author[ISKP]{ F.~Mosel} 
\author[HH]{ M.~M\"uller} 
\author[HH]{ M.~M\"unstermann} 
\author[IKP]{ D.~Prasuhn}
\author[IKP]{ P.~von~Rossen}
\author[ISKP]{ H.~Scheid} 
\author[HH]{ N.~Schirm} 
\author[ISKP]{ F.~Schwandt} 
\author[HH]{ W.~Scobel} 
\author[ISKP]{ H.~J.~Trelle} 
\author[HH]{ A.~Wellinghausen} 
\author[ISKP]{ W.~Wiedmann} 
\author[HH]{ K.~Woller} 
\author[ISKP]{ R.~Ziegler} 
\corauth[cor1]{Heiko Rohdjess, FAX: +49-228-73 2505}
\address[ISKP]{Helmholtz-Institut f\"ur Strahlen- und Kernphysik, Universit\"at Bonn,
Germany}
\address[HH]{Institut f\"ur Experimentalphysik, Universit\"at Hamburg,
Germany}
\address[IKP]{Institut f\"ur Kernphysik, Forschungszentrum J\"ulich, Germany}

\begin{abstract}
The EDDA-Detector at the Cooler-Synchrotron COSY/J\"ulich has been
operated with an internal \CH\ fiber target to measure proton-proton
elastic scattering differential cross sections. For the data analysis
knowledge of beam parameters, like position, width and angle, are
indispensable. We have developed a method to obtain these values with
high precision from the azimuthal and polar angles of the ejectiles only, by
exploiting the coplanarity of the two final state protons with the
beam and the kinematic correlation. The
formalism is described and results for beam parameters 
obtained during beam acceleration are given.
\end{abstract}

\end{frontmatter}

\section{Introduction}
In nuclear and particle physics experiment knowledge of the
distribution of scattering vertices is a crucial ingredient of data
analysis. In most cases experiments are designed to yield enough
position information to trace detected particles back to the origin.
However, even with a much simpler detector, events with particularly
simple topologies like elastic scattering may allow to extract the
vertex distribution with much lesser experimental information.

\begin{figure}
\includegraphics[width=\mytextwidth]{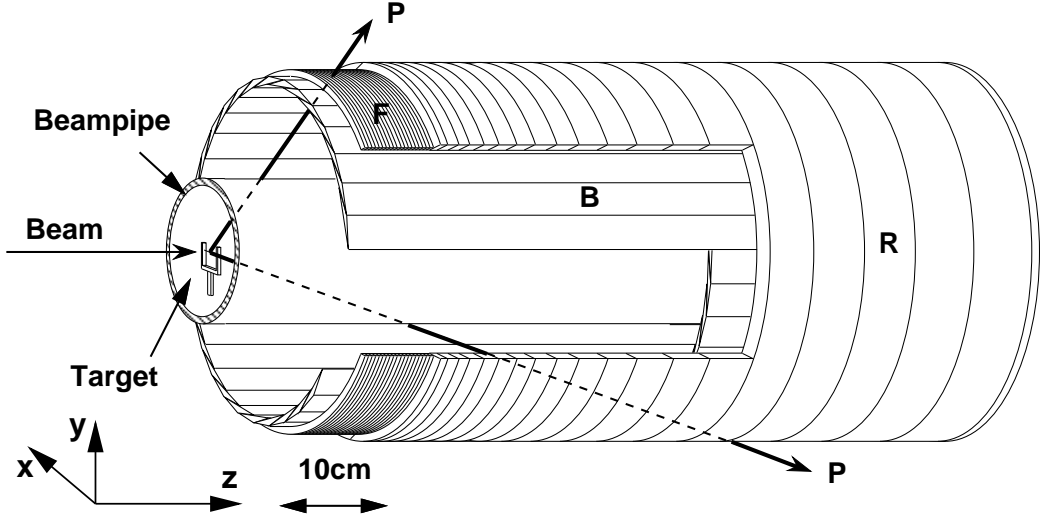}
\caption{Schematic drawing of the EDDA detector}
\label{fig:det}
\end{figure}

The EDDA experiment \cite{albers97,Altmeier:2004} at the COSY
accelerator \cite{Maier:1997} in J\"ulich is dedicated to
the measurement of pp elastic scattering over a wide
energy (0.5-2.5~GeV) and angular range (30\gradx\ to 90\gradx\ in the
c.m.). The experimental program includes measurements of the
unpolarized differential cross-section \cite{albers97,Altmeier:2004},
analyzing power \cite{Altmeier:2000} and spin-correlation parameter
\cite{Bauer:2002zm}. Data taking for unpolarized cross section was
done only with the outer part of the EDDA-detector, shown in
\Fig{fig:det}, detecting in coincidence a single point of incidence
for each of the two protons, originating from elastic scattering of
the internal proton beam from $5\times4\ \mu$m$^2$ thin \CH-fiber targets. 
However, for the reconstruction of scattering angles, detector
calibration, and luminosity determination detailed knowledge of the 
beam-target overlap is important. Since measurements were taken during
acceleration of the COSY-beam, such that beam position and width are a
function of beam-momentum, it was mandatory, that these quantities can
be determined from the scattering data itself.

Two-body kinematics require the two protons to be coplanar with the
beam. Thus deviations from coplanarity of proton-proton scattering
events with the detector symmetry axis bears information on the
position and angle of the COSY beam axis with respect to the detector.
Along these lines we developed a method to extract information on beam and
target parameters like width, position and angle from a single-layered
detector described in this paper. In sections 2 and 3 we will give a
brief description of the experiment and the selection of scattering
data. Section 4 will explain the formalism and in section 5
some examples for the application of this method will be discussed.

\section{Experiment}
The EDDA Detector (cf. \Fig{fig:det}) measures
the point of interception of the two final-state particles of
proton-proton scattering events with a cylindrical symmetric
scintillator array around the COSY beam pipe. It consists of a layer
of 32 overlapping scintillator bars (B) of triangular cross section
detecting the azimuthal angle $\phi$ and a layer of 
2$\times$29 semirings, made either from bulk scintillator (R) or scintillating
fibers (F), detecting the position $z$ of incidence along the symmetry axis
of the detector and bears information on the polar angle \Thlabx.
The angular resolution is improved by about a factor of five with
respect to the granularity by comparing off-line the light output of two
adjacent detector elements  \cite{bisplinghoff93}.

Thin \CH- or carbon-fibers, stretched horizontally on a 30~mm wide
fork, are used 
as targets. They can be moved vertically in and out of the beam by a
magnetic drive. Beam lifetimes of 3 to 10~s, depending on beam energy,
have been achieved. Data is taken during and/or after acceleration of
the COSY-beam and accumulated over many machine cycles to obtain
statistical precision. Hits in opposing bars serve as a coplanarity
trigger.
In addition, elastic scattering kinematics
\begin{equation}
\cot\Thlabi{1}\cdot\cot\Thlabi{2} = \frac{z_1\cdot z_2}{R_R^2} = \gcm^2,
\label{eq:kin}
\end{equation}
with \gcm\ being the Lorentz-factor of the c.m. in the laboratory and
$R_R$ the radius of the ring-layer, is tested 
by a coincidence of two semiring with appropriate locations in $z$.

\section{Event Selection}

\begin{figure}
\begin{center}
\includegraphics[width=10cm]{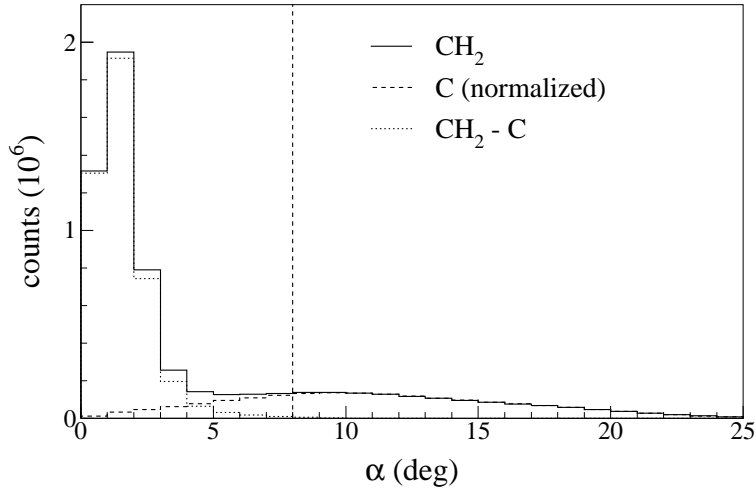}
\end{center}
\caption{Distribution of the kinematic deficit $\alpha$ for scattering
events at a beam momentum of 1.455~GeV/c. The upper limit for accepted
elastic scattering 
events is shown as the vertical line.}
\label{fig:alpha}
\end{figure}

Coplanarity and the kinematic correlation (cf. \Eq{eq:kin}) can be
tested with refined precision offline. Then two prongs detected in
coincidence are transformed to the c.m. system, with the hypothesis of
being elastically scattered protons. For pp elastic scattering events
these prongs should be back-to-back in the c.m. system, and the
angular deviation $\alpha$ from this signature, the so-called
kinematic deficit, should be small. A typical distribution is shown in 
\Fig{fig:alpha}. The apparent contribution from the carbon contents of
the target is measured separately using pure C-targets and subtracted
statistically. To this end two data sets taken with a \CH- and a
C-target under the same experimental conditions are subject to an 
identical offline analysis. The relative normalization of the two
samples with respect to proton-carbon reactions is derived from the
counts in the range 10\grad to 15\gradx in $\alpha$
(cf. \Fig{fig:alpha}), mainly populated by quasi-free scattering on
protons bound in carbon. This normalization factor is used to subtract
the contribution from pC scattering in all acquired experimental
spectra, so that a clean pp elastic scattering sample is
available. Monte-Carlo simulations showed this procedure also takes
care of the small contribution from inelastic pp-reactions in first
order. Details of the event reconstruction are described elsewhere 
\cite{albers97,Altmeier:2004}. 

\section{Formalism}

In this section we will use a coordinate system ($x,y,z$) 
attached to the EDDA detector as shown in \Fig{fig:det}, with the
$z$-axis being the symmetry axis of the detector pointing downstream
from the target (at $z=0$) with respect to the COSY-beam, the $y$-axis
pointing upward in the 
lab, and the $x$-axis pointing outboard of the COSY-ring. 
If not stated otherwise all quantities will be given in this detector
coordinate system. Widths given as standard deviation are denoted by
$\sigma$, while $\delta$ signifies FWHM (full width at half maximum) 

The distribution of scattering vertices must be along the fiber
target, which have negligible (4-5~$\mu$m) spatial extension in the
$y$- and $z$-directions. The longitudinal target position along $z$ can
be determined from scattering data using \Eq{eq:kin} and defines the
zero of the coordinate system in this direction. Finally, the
parameters yet to be determined are 
\begin{tabbing}
  xxxxx\= xxxxxxxxxxxxx\= \+\kill
\yb\ \> fiber target position in $y$ at $z$=0. \\
\xb\ \> beam position in $x$ at $z$=0 and \yb . \\
\xp\ \> horizontal beam angle (d$x$/d$s$) at $z$=0. \\
\yp\ \> vertical beam angle (d$y$/d$s$) at $z$=0. \\
\sxb\ \> size of the beam in $x$ at $z$=0 and \yb .\\
\delp\ \> detector resolution in the azimuthal angles $\phi_1$ and $\phi_2$.  
\end{tabbing}

For simplicity the formalism will first be described for a beam
parallel to the detector symmetry axis, i.e. \xp = \yp = 0. The
generalization to tilted beams will be the topic of Section~\ref{sec12}

\subsection{Beam Parallel to the Detector Axis}
\label{parallel}

\begin{figure}
    \begin{center}
      \includegraphics[width=9cm]{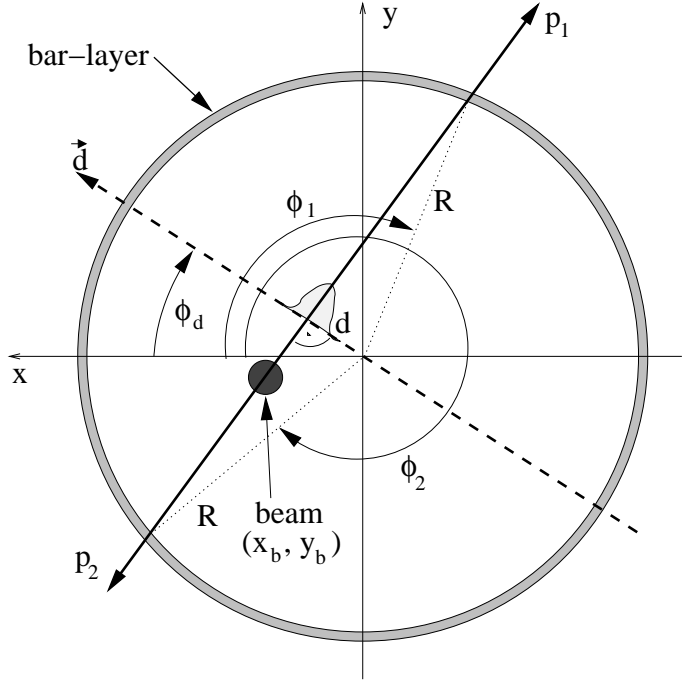}
    \end{center}
  \caption{Geometry for a beam parallel to the detector symmetry axis $z$ as explained in the text.}
  \label{geo}
\end{figure}

Consider a beam displaced by (\xb ,\yb ) in the $x$,$y$-plane from the
detector symmetry axis. In a projection on the $x$,$y$-plane of an
elastic scattering event (prongs $p_1$ and $p_2$ in \Fig{geo}) a beam
displacement leads to a deviation from coplanarity when the vertex
{\em is assumed} to be at ($x$,$y$)=(0,0) (dotted lines). 

The EDDA detector measures the point of interception of these two
prongs with the bar layer at a mean radius of $R=164$~mm. Connecting
these two points by a straight line its distance $d$ to the origin of the
coordinate system is a function of the beam position. If we now define
an axis $\vec{d}$ perpendicular to this line (fat dashed line in
Fig.~\ref{geo}) we can 
calculate the angle \ddp\ of the $\vec{d}$ axis to the $x$-axis
\begin{equation}
  \ddp = \frac{\phi_1 + \phi_2}{2}
  \label{phiddef}
\end{equation}
and the position $d$ along this axis:
\begin{equation}
  d = - R \cos\frac{\phi_1 - \phi_2}{2}
  \label{ddef}
\end{equation}
The orientation of the  $\vec{d}$ axis has been chosen to point
to the ``left'', to the effect that \mbox{$\ddp\in[-\pi/2,\pi/2]$} must be
forced by adding or subtracting $\pi$ from the expression given in
\Eq{phiddef} (For the example of \Fig{geo} $\pi$ must be subtracted). The negative sign in  \Eq{ddef} is due to this
choice. \Eqs{phiddef}{ddef} are an eventwise
transformation from the observables ($\phi_1$,$\phi_2$) to ($d$,$\phi_d$). 

The connection between $d$ and the beam position is evident. Using
basic trigonometric relations (for a derivation see Appendix A)  -- or
by a simple geometric derivation using \Fig{geo} -- one
obtains:
\begin{equation}
d = \xb \cos\ddp + \yb \sin\ddp
  \label{dscalesimple}
\end{equation}
Now, the idea is to consider the $d$-distribution of events with
different \ddp\ in order to deduce \xb\ and \yb. 

\subsubsection{Experimental Input and Analysis}
\begin{figure}
    \begin{center}
      \includegraphics[width=13cm]{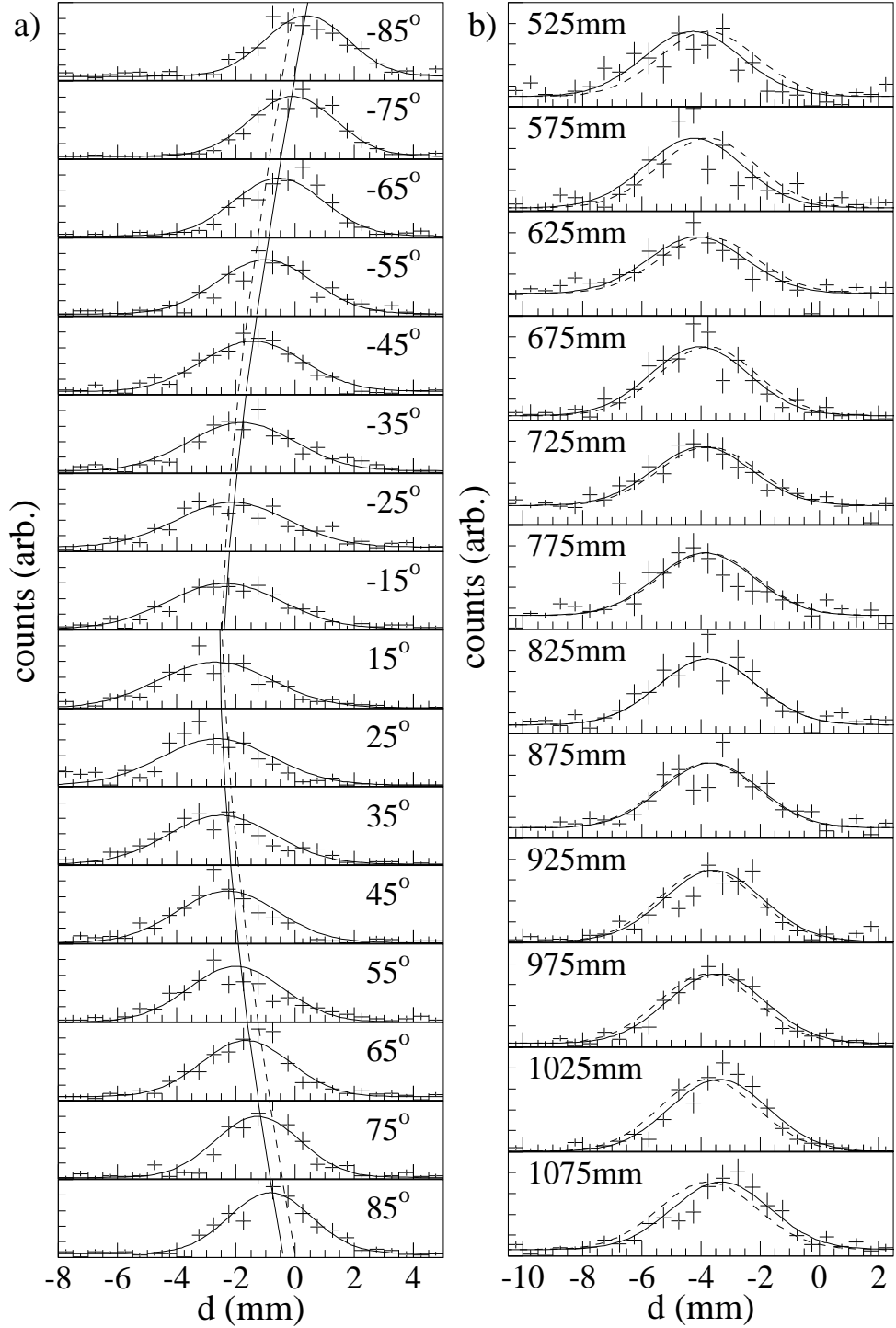}
    \end{center}
  \caption{(a) Variation of the $d$-distribution for different bins in
      \ddp. The fit (solid Gaussian) to the data (crosses) yields a
      beam displacement of $\xb = -2.51$~mm, $\yb = -0.41$~mm. The
      mean value is shown as the solid vertical curve, for the dashed
      curve \yb\ was fixed to zero.
      (b) Variation of the $d$-distribution at $\ddp = 15\pm5\gradx$
      with $z_1+z_2$ and the result of the fit (solid line) yielding
      $\xp~\approx5$~mrad. The dashed line shows the fit at 825~mm for
      comparison. 	
    }
  \label{fig:xxprime}
\end{figure}

For each event \ddp\ and $d$ are calculated. The events are sorted
into spectra vs. $d$ for different ranges in \ddp. Events with $|\ddp
|<10^\circ$ are discarded, since here the readout of the semirings
causes trigger inefficiencies.
A $d$-distribution is accumulated for 16 \ddp -intervals from
-90$^\circ$ to -10$^\circ$ and 10$^\circ$ to 90$^\circ$, each
5$^\circ$ wide. These 16 spectra, an example is shown in
\Fig{fig:xxprime}~(a), are the input to the analysis.
For investigation of the time dependence of beam parameters this set of
spectra is accumulated for small time intervals, usually 25, 50 or
100~ms wide, separately.

The experimental $d$-distributions  can be
fitted by a Gaussian 
\begin{equation}
  n(d) = \frac{N(\ddp)}{\sqrt{2\pi}\dds } \exp\left(-\frac{(d-\ddb
    )^2}{\dds^2}\right) 
  \label{gauss}
\end{equation}
with two fit parameters \ddb\ and \dds\ bearing the information on the
beam. The total number of counts for
a given \ddp-interval $N(\ddp)$ is used for the normalization and kept
fixed in the fit.

For a given \ddp, the value of \ddb\ is given by \Eq{dscalesimple} and
\begin{equation}
\dds \approx \sqrt{\cos^2\ddp \sigma^2_{\xb} + \frac{R^2}{2} \sigma^2_\phi }
  \label{dsfit}
\end{equation}
when the vertical size of the target is neglected (cf. Appendix A).
Thus all 16 spectra are fitted {\em simultaneously} by four fit
parameters:  \xb , \yb, $\sigma^2_{\xb}$, and $\sigma^2_\phi$. 
For the implementation of the fitting procedure the minimizer MIGRAD
(variable metric method) of the CERN package MINUIT
\cite{james94} is used.
Typical spectra are shown in \Fig{fig:xxprime} (a) for a 25~ms wide
bin in time -- corresponding to $\Delta p \approx 29$~MeV/c during
acceleration of the COSY beam. Clearly, differences of 0.5~mm in \yb\
can be resolved 
by the data.

\subsubsection{Test of Method}

\begin{figure}
  \begin{center}
    \includegraphics[width=12cm]{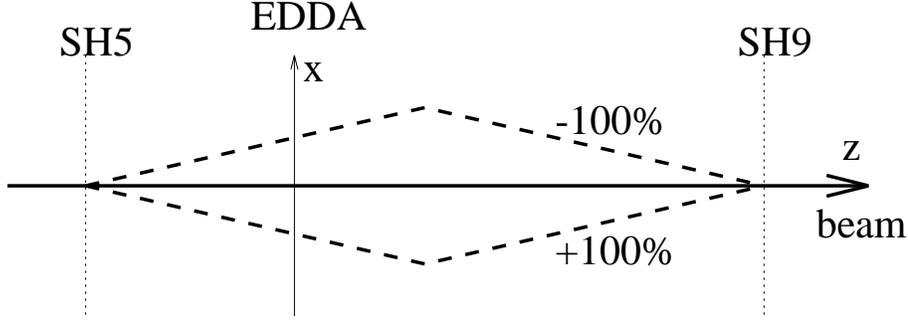}
  \end{center}
  \caption{Sketch of the orbit distortion expected for a simultaneous
    excitation of the horizontal steerer SH5 and SH9. The beam
    position is measured by EDDA at the indicated position.
    }  
  \label{steerer}
\end{figure}

\begin{figure}
    \begin{center}
      \includegraphics[width=\mytextwidth]{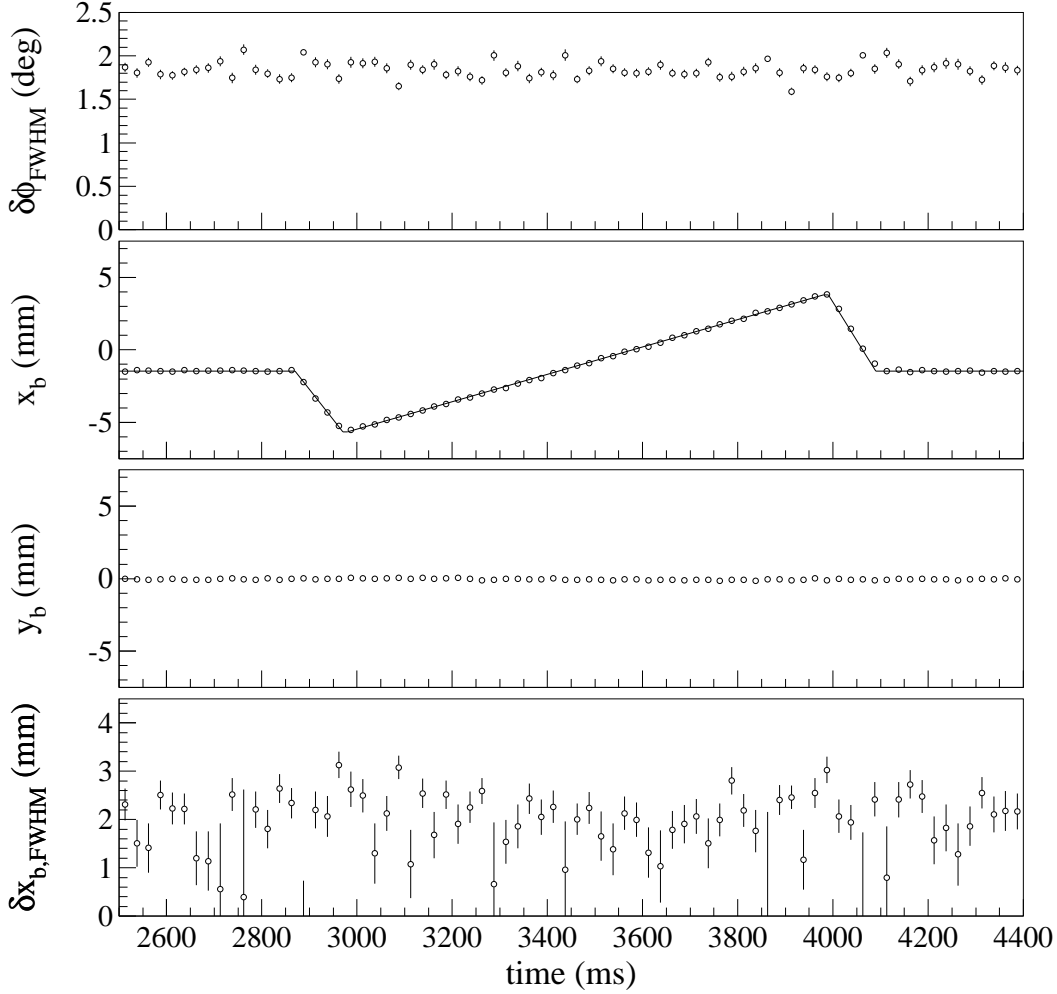}
    \end{center}
  \caption{Result of the fit for flattop data at 3.4~GeV/c
    where around 3000-4000~ms a horizontal excursion of the beam
    was produced by exciting steerer magnets (cf. Fig.~\protect\ref{steerer}).
    The solid line shows the expected shape of the excursion with the
      amplitude and position in \xb\ adjusted to the data.
    }
  \label{str9first}
\end{figure}

\begin{figure}
  \begin{center}
    \includegraphics[width=\mytextwidth]{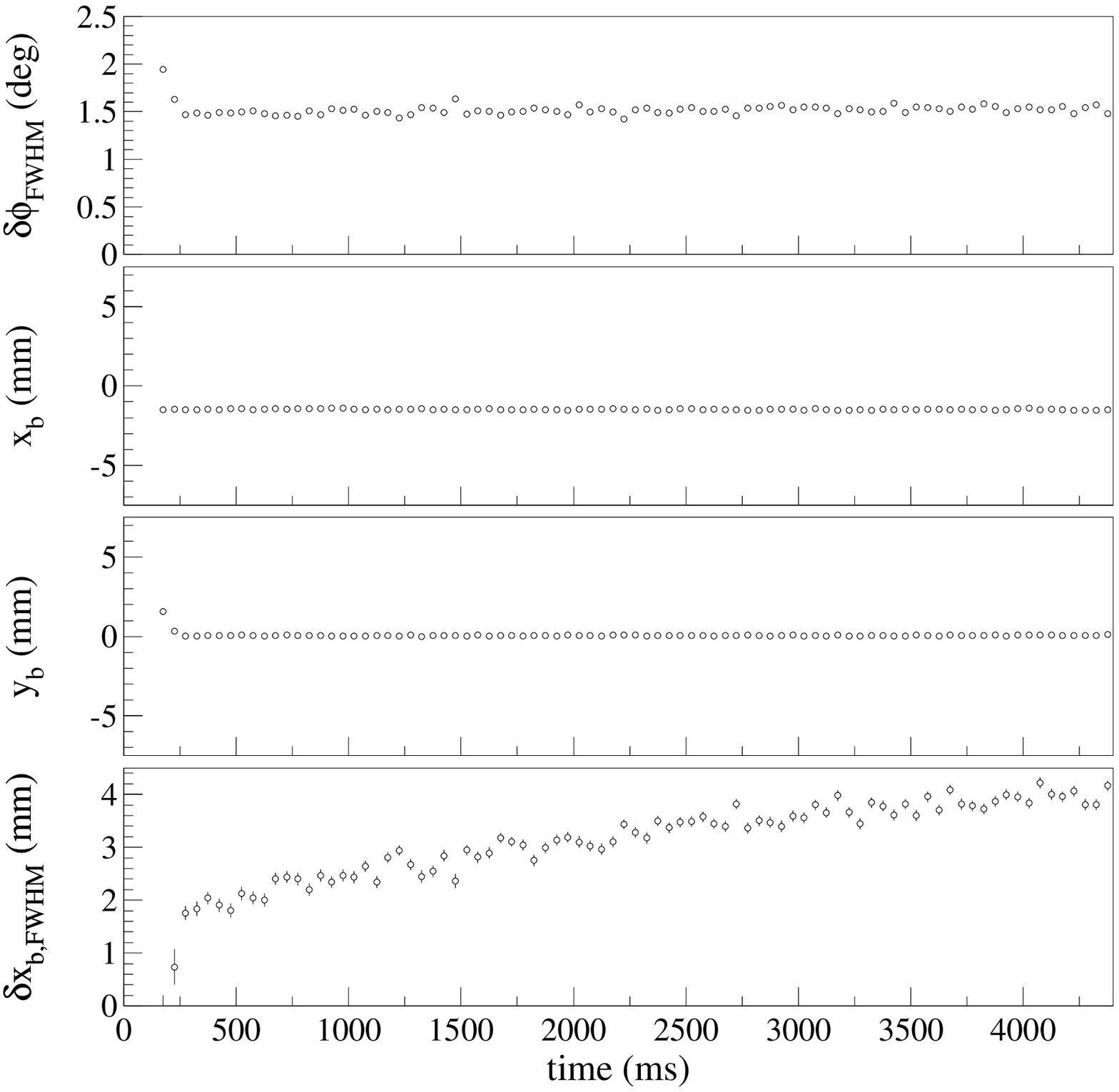}
  \end{center}
  \caption{Result of the fit for flattop data at 2.25~GeV/c. 
	 The fiber target starts to move at 0~ms and reaches the beam
    position around 250~ms.
    }
  \label{str9firstflat}
\end{figure}

As a test of the sensitivity of the method, data at fixed momentum of
3.4~GeV/c was acquired while the beam was moved horizontally across the 
target by exciting two horizontal steerer dipoles of the COSY-lattice.
Between the steerer, separated by a phase advance of about $\pi$, 
focusing quadrupoles are located, resulting in the geometry of the beam
excursion as depicted in Fig.~\ref{steerer}.
The strength of the magnets
(originally at 0\%) was ramped within 100~ms to +80\%, then, after
10~ms, they were ramped linearly to -100\% within 1000~ms and finally
--after another 10~ms-- brought back to 0\% within 100~ms.

The result of the fit (Fig.~\ref{str9first}) shows the response of the
beam: The beam position \xb\ follows closely the strength of the
magnet.  The vertical target position \yb , the beam width \sxb ,
and the angular resolution \delp\ remain basically unchanged.
This proves the sensitivity of this method to the beam position.

Another test is to take data taken at fixed energy and beam position
and to observe the increase in beam size due to the emittance growth
when the beam is ``heated'' by the target. The result of the fit is shown in
Fig.~\ref{str9firstflat}. As expected only the beam size increases
with time. All other parameters remain constant. It is interesting to
see, that even the overshoot of the target -- after is has been moved in
vertically during the first 250~ms by a linear actuator and before it
settles at the 
correct height -- is resolved. As long as the target moves the vertex
is then spread out in $y$ and results in an increased angular
width \delp\ in the fit.

In all figures only statistical errors are given as they are
calculated from the error-matrix of the nonlinear \chiz-fit.
However, systematic errors arise mainly due to strong correlations of
fit parameters. In \Fig{str9firstflat} the anti-correlations of \delp\ and
\sxb\ introduced by \Eq{dsfit} is clearly visible. Since 
$R\delp/\sqrt{2} \approx 3.6$~mm the angular resolution in $\phi$ makes
it difficult to determine beam sizes smaller than 3~mm accurately.
Then, the $\cos^2\ddp$ modulation of \dds\ becomes very small and
statistical fluctuations start to mislead the fit.

\begin{figure}
    \begin{center}
      \includegraphics[width=\mytextwidth]{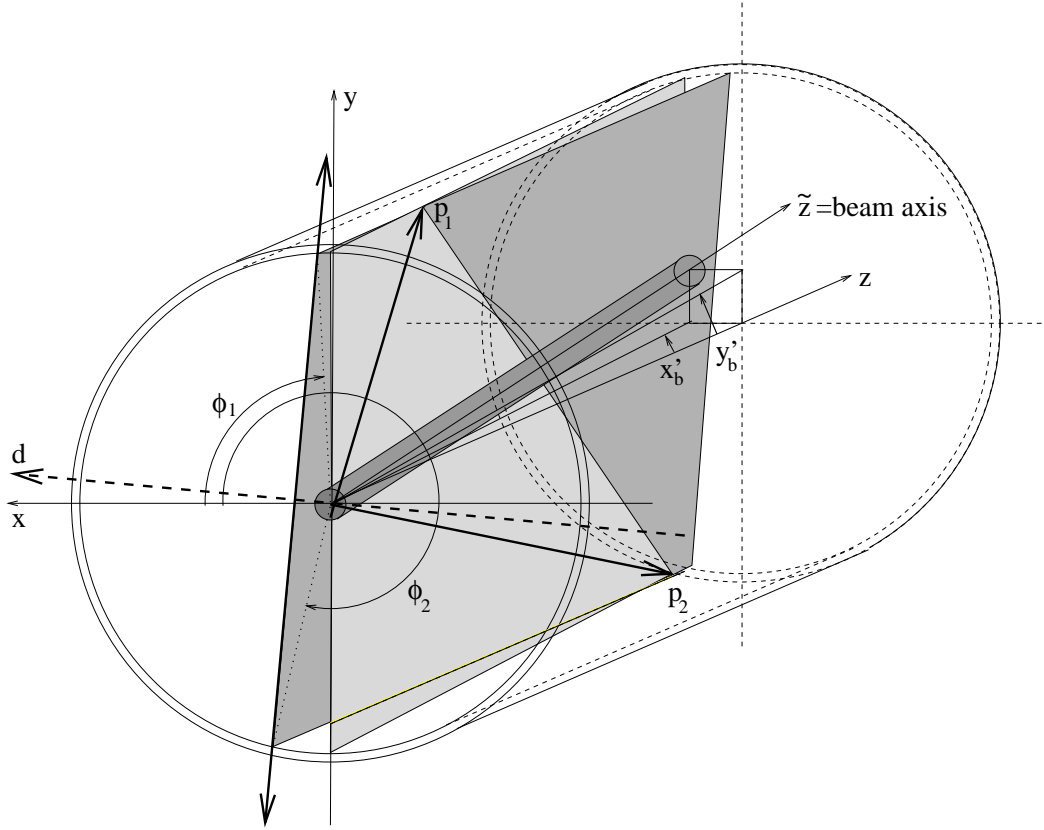}
    \end{center}
  \caption{Geometry for a tilted beam (axis \zt\ tilted by \xp , \yp )
    as explained in the text.}
  \label{geo2}
\end{figure}

\subsection{Tilted Beam}
\label{sec12}
In the previous section it has been assumed that the beam is parallel
to the symmetry axis of the detector. This is not necessarily true.
In general, the beam may be tilted with respect to the detector both
horizontally and vertically, described by the parameters \xp\  and \yp
.
Fig.~\ref{geo2} shows that a tilted, non-displaced beam mocks up a
beam displacement in the analysis neglecting \xp\ and \yp: since only
the azimuthal 
angles are considered, we reconstruct the event in a plane parallel to
the detector z-axis, defined by the two points of interception with the
EDDA detector (dark shaded plane in Fig.~\ref{geo2}). The reaction plane
(light plane in Fig.~\ref{geo2}) is defined by the same two points but
must be coplanar to the beam. Consider a pp-elastic event where the
two protons are emitted up-down: if the beam angle \xp\ is positive
the two protons tracks will be displaced to the left with respect to
their origin. The displacement depends on the distance in $z$ they
travel until they hit the EDDA bar layer at radius $R$. As evinced in
Fig.~\ref{geo2} the line through the two points of interception
projected into the $x$-$y$ plane looks as if the beam were displaced
to positive $x$.

In Appendix A it is derived that the measured value of $d$ of a tilted
beam is given by:
\begin{equation}
  d = (\xb + \frac{z_1+z_2}{2} \xp) \cos\ddp + (\yb + \frac{z_1+z_2}{2} \yp) \sin\ddp
  \label{tilted}
\end{equation}
and reduces to \Eq{dscalesimple} for $\xp = \yp = 0$.
$z_1$ and $z_2$ are the $z$-coordinate of the point where the tracks
of the proton intercept the EDDA bar layer. 
Evidently, the parameters \xb\ and \xp\ (as \yb\ and \yp ) are
strongly correlated and can only be distinguished by considering
events with different $z_1+z_2$.
For pp-elastic scattering the correlation of the two $z$ values as a
function of beam kinetic energy $T_p$ is
given by \Eq{eq:kin} 
\begin{equation}
  z_1 z_2 = R^2\gcm^2 = R^2 (1 + \frac{T_p}{2m_p} )
  \label{zcorr}
\end{equation}
where $m_p$ is the proton mass.
The values covered by $z_1 + z_2$ range from $2 R\gcm \approx
400\ldots500$~mm for symmetric events ($z_1 = z_2$) to
$1100$~mm for asymmetric events ($z_1 = z_{\rm max} = 1$~m, see \Fig{fig:det}).

\begin{figure}
    \begin{center}
      \includegraphics[width=\mytextwidth]{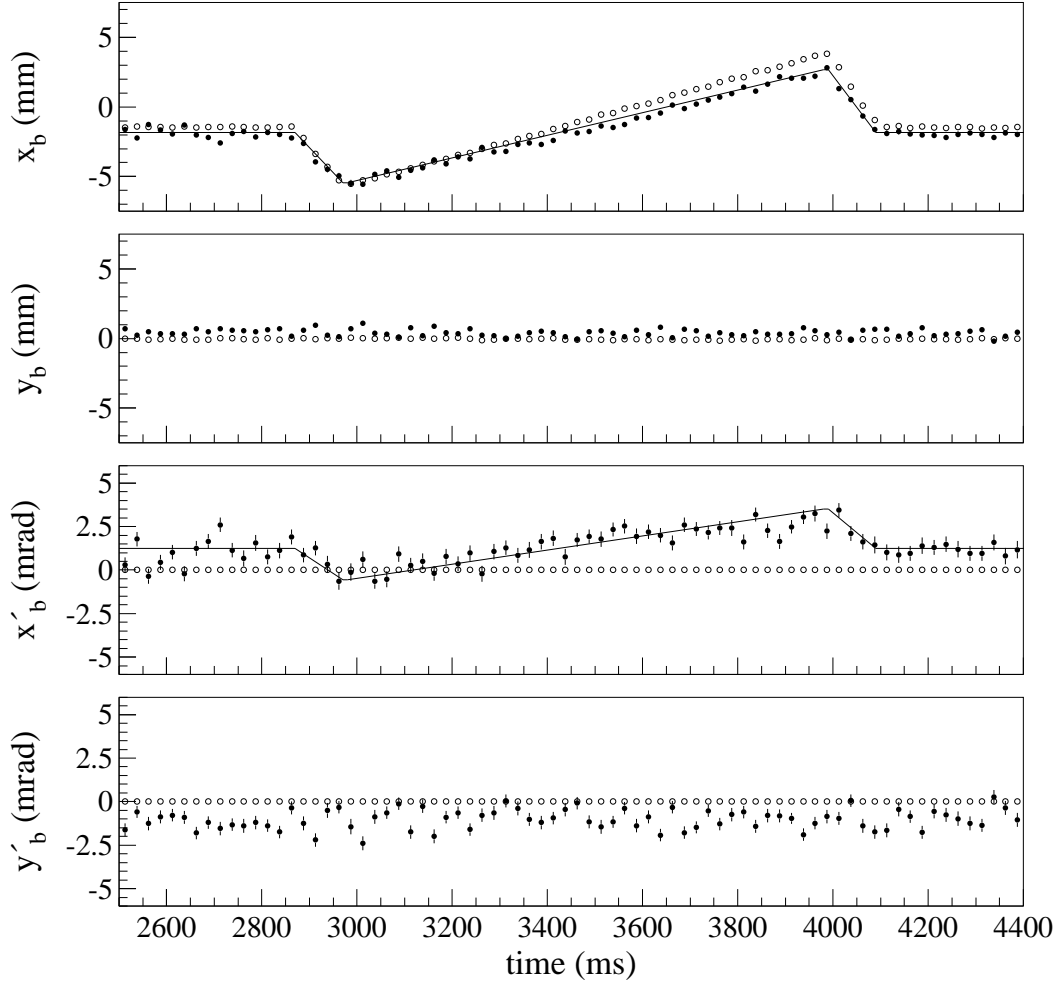}
    \end{center}
  \caption{Result of the fit for flattop data at 3.4~GeV/c. The solid symbols
    are the result of the fit when the beam angles \xp\ and \yp\ are
    included in the fit. The open symbols are the data as in
    Fig.~\protect\ref{str9first} and the solid lines the expected
      shape of the excursion. 
    }
  \label{fig:str9tilt}
\end{figure}

\subsubsection{Experimental Input}
For each \ddp-bin  the data was further subdivided into 14
classes with respect to their $z_1+z_2$ values, ranging from
400 to 1100~mm, each 
50~mm wide. For a given energy only those classes were included in the 
fit where the kinematics allowed for the respective $z_1+z_2$
value. Usually 8-13 classes are useful, and for each class 16 $d$
distributions for the different \ddp\ ranges are sorted. 
The resulting 128-208 spectra are fitted {\em simultaneously} with six
parameters \xb , \yb , $\sigma_\phi$, $\sigma_{\xb}$, \xp , and \yp.
An example of these spectra is shown in \Fig{fig:xxprime}~(b) where
the shift in \ddb\ with $z_1 + z_2$, containing the information on
beam angles is clearly visible through comparison to the result for
$z_1+z_2\approx 825$~mm.

The result of the
fit (same data as in Fig.~\ref{str9first}) is shown as the solid
symbols in Fig.~\ref{fig:str9tilt}. The 
change in beam angle and the beam position have the same sign as one
expects for the geometry shown in Fig.~\ref{steerer}.
The total change in angle during the bump (4~mrad) and the range of
the beam position shift (8~mm) yield a distance of the origin of the
beam rotation at -2~m with respect to the target location which agrees
rather well with 
the distance of the steerer magnet SH5 to the fiber target (1.75~m).
The fact that the change in \xb\ and \xp\ have the same sign indicates
that the method is indeed sensitive to changes in beam angles, since
\Eq{tilted} implies that fluctuations of these parameters due to
statistics would be {\em  anti-correlated} in the fit. 
Again errors shown are purely
statistical, and the observed fluctuations around the expected
excursions are now much larger, due to the strong anti-correlation of
beam position and angle in the fit. These fluctuations can be only
be damped by improving on the statistical precision of the data.
Summarizing, possible beam tilts \xp\ and \yp\ should always be taken
into account since the results for the beam positions \xb\ and \yb\
depend on \xp\ and \yp\ (compare \Figs{str9first} and \ref{fig:str9tilt}).
\section{Results and Discussion}
\begin{figure}
    \begin{center}
      \includegraphics[width=\mytextwidth]{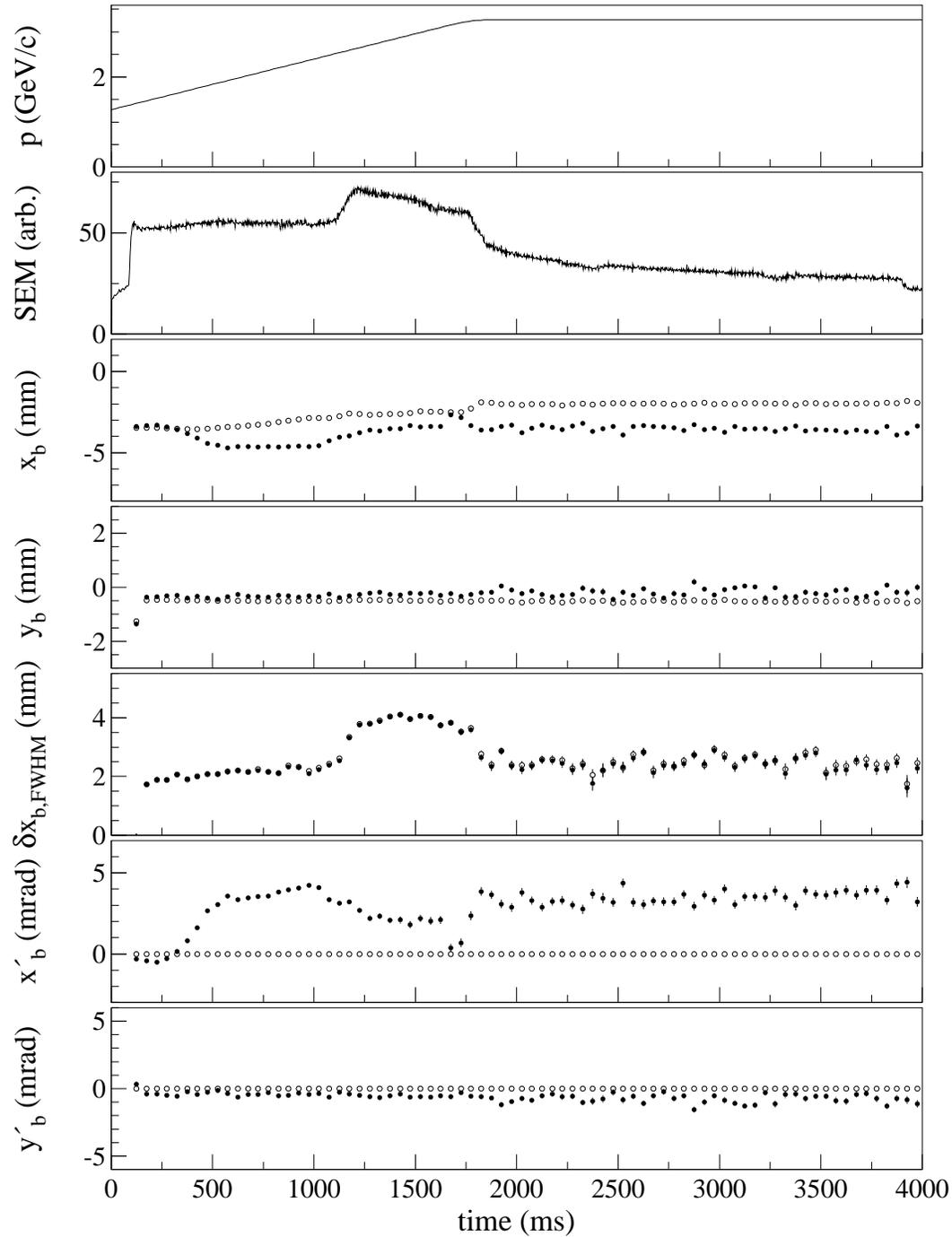}
    \end{center}
  \caption{Fit parameter as a function of time for data acquired during
      acceleration and the flattop. Also shown is the momentum of the
      beam and the signal of a luminosity monitor, the secondary
      electron monitor (SEM). Solid (open) symbols and
    solid (dashed) lines show the
    result of the fit when \xp\ and \yp\ were (were not) included as
    fit parameters.}
  \label{fig:str7ramp}
\end{figure}

The method of reconstructing the first and second moment of the
vertex distribution has been applied routinely for data analysis of
the EDDA experiment. In \Fig{fig:str7ramp} we show an example of a
measurement taken during acceleration of the COSY beam and the flattop
at 3.3~GeV/c. For comparison both the beam momentum and the response of the
secondary electron monitor (SEM) as a measure of the luminosity are
shown. In this example the beam position could be held rather
stable throughout the 
ramp, some shifts in the beam angle \xp\ is seen, especially in the
transition to the flattop. Most strikingly, the rise in the SEM signal
coincides with a dramatic increase in the horizontal beam width. Beam
profile measurements showed a correlated {\em decrease} of the
vertical beam width. Thus, the increase in luminosity on the
horizontally mounted EDDA fiber target is due to a change of shape of
the beam ellipse in the x-y plane at the target location, probably
caused by a resonant coupling 
of the vertical and horizontal phase-space.
The results show that including the beam angles in the fit is
very important to reliably extract the beam position, besides the \chiz\
per degree of freedom is considerably reduced.
The achieved resolution is better than 0.5~mm for beam position and width and
about 0.5~mrad for beam angles. 

Statistics obtained at typical running conditions of the EDDA
experiment are sufficient to obtain this information with 25~ms
resolution, corresponding to less than 30~MeV/c in momentum bins during
acceleration with ${\rm d}p/{\rm d}t = 1.15$~(GeV/c)/s ramping speed.
Limitation of this method are, that displacements ($<$10~mm) and beam
angles ($<$10~mrad) must be sufficiently small, as it was routinely
achieved at COSY. Furthermore, beam widths below 2.5~mm
FWHM cannot be determined very accurately, due to the limited
azimuthal resolution.

\section{Summary}
Exploiting the fact that pp-elastic events are coplanar, a measurement
of the position where the particles intercept a cylindrical
scintillator hodoscope, yields information on the position, width and
angle of the impinging beam, without individual tracing of the
particle tracks.  
The horizontal beam position and the vertical target position can be
obtained with an accuracy of about 1~mm, beam angles to about 1~mrad and
the beam width to better than 1~mm FWHM (provided the beam is at least
2.5~mm wide). This information is obtained for time intervals of 25~ms
which is sufficient to follow the smooth drift of a stored, accelerated 
proton beam in a synchrotron like COSY.

We gratefully acknowledge excellent beam preparation by the COSY
operating staff, including the setup of the beam excursions used for
consistency tests.

\appendix
\section{Derivation of the Formalism}
\subsection{Coordinate Systems}

When the beam is both displaced and tilted with respect to the
detector it is convenient to introduce a ``beam coordinate
system'' BC (\xt,\yt,\zt) 
where the reaction vertex is at (0,0,0) and the beam
parallel to the \zt-axis. Transformations from BC to
the ``detector coordinate system'' DC ($x,y,z$) are then described by a
rotation by an 
orthogonal matrix \mmm\ followed by a translation to the vertex
position ($x_b$,$y_b$,0) (we assume the vertex is at z=0)

\begin{equation}
\left( \begin{array}{c}x \\ y \\ z  \end{array}\right) = 
\left( \begin{array}{c}x_b \\ y_b \\ 0 \end{array} \right) +
\mmm \left( \begin{array}{c}\xt \\ \yt \\ \zt\\ \end{array}\right) 
  \label{mat2}
\end{equation}

The transformation matrix can be derived by expressing the unit
vectors $e_{\xt}$, $e_{\yt}$, and $e_{\zt}$ of the BC system in the DC
space for the special case $x_b$=$y_b$=0:
\begin{itemize}
\item the unit vector in \zt\ is given by the beam parameter $x'_b = {\rm
    d}x/{\rm d}s$ and $y'_b = {\rm d}y/{\rm d}s$. If one defines angles
  $\theta_x$ and $\theta_y$ via $\tan\theta_x = x'$ and $\tan\theta_y
  = y'$ one obtains $e_{\zt } = (\sin\theta_x,\sin\theta_y,k)$, where k is
  fixed by normalization
  \begin{equation}
    k=\sqrt{1-\sin^2\theta_x-\sin^2\theta_y}.
    \label{k}
  \end{equation}
\item for convenience we choose $e_{\xt}$ to be in the x-z
  plane. Orthogonality to $e_{\zt} $ and unit length fix $e_{\xt}$ to be
  $(k/\cos\theta_y,0,-\sin\theta_x/\cos\theta_y)$.
\item $e_{\yt} $ is then given by the cross product $e_{\zt } \times e_{\xt}$.
\end{itemize}
The transformation matrix is thus given by:
\begin{equation}
\mmm = \left( 
\begin{array}{ccc}
    k/\cos\theta_y             & -\sin\theta_x\tan\theta_y  &  \sin\theta_x \\
    0                          &     \cos\theta_y           &  \sin\theta_y \\
    -\sin\theta_x/\cos\theta_y & -k \tan\theta_y            &    k \\
\end{array}    
\right)
\label{matrix}
\end{equation}
The inverse transformation is obtained by inverting \mmm . Since \mmm\
is orthogonal we simply use the transposed matrix:
\begin{equation}
\left( \begin{array}{c}\xt \\ \yt \\ \zt\\ \end{array} \right) = 
\mmm^T \left( \begin{array}{c}x - x_b \\ y - y_b \\ z  \end{array}\right) 
  \label{mat1}
\end{equation}

\subsection{Dependence of \ddb\ on Beam Parameters}
The quantity \ddb\ (as explained in Section 1) is given by
\begin{equation}
\ddb\ = - R \cos{\frac{\phi_1-\phi_2}{2}}
  \label{defd}
\end{equation}
as the {\em average} $d$ for many events, so that smearing due to the
finite detector resolution is averaged out.

Lets now consider a perfect proton-proton scattering event
with azimuthal angles
$\tilde{\phi_i}$ and polar angles $\tilde{\theta_i}$ in the BC system.
The unit vectors of their direction are given by
\begin{equation}
  \vec{\tilde{p_i}}  =  
  \left(\begin{array}{c}\sin\tilde{\theta_i} \cos\tilde{\phi_i} \\ 
        \sin\tilde{\theta_i}\sin\tilde{\phi_i} \\
    \cos\tilde{\theta_i}\\ 
  \end{array}\right) \ \ \ \ \ \ \ \ ; \ \ i=1,2
  \label{pptilde}
\end{equation}

We now calculate the point of interception of the two prongs
with the cylindrical
bar layer of the EDDA detector at radius R in the DC  system viz
\begin{equation}
  \left(\begin{array}{c} R\cos\phi_i \\ R \sin\phi_i \\ z_i \\
    \end{array}\right) 
    = \left( \begin{array}{c}x_b \\ y_b \\ 0 \end{array} \right) +
    \lambda_i\ \mmm\  \vec{\tilde{p_i}} \ \ \ \ \ \ \ \ ; \ \ i=1,2
  \label{inter}
\end{equation}
by determining the three parameters $z_i$, $\phi_i$ (describing the
point where the bar layer is hit) and $\lambda_i$.
The result can not be obtained analytically, however, for some
selected values for $x_b$, $y_b$, $x'_b$, and $y'_b$ the general
dependence can be extracted.
Therefore, we first approximate the matrix \mmm\ for small angles 
$x'_b$ and $y'_b$ with the result
\begin{equation}
\mmm = \left( 
\begin{array}{ccc}
    \sqrt{\frac{1-\xps - \yps}{1-\yps}} & 
    -\frac{\xp \yp}{\sqrt{1-\yps}} & \xp  \\
     0 & \sqrt{1-\yps} &  \yp  \\
    -\frac{\xp}{\sqrt{1-\yps}} & 
    -\yp \sqrt{\frac{1-\xps - \yps}{1-\yps}}  &  
    \sqrt{1- \xps - \yps} \\
\end{array}    
\right)
\label{matrix2}
\end{equation}
and consider the following cases:

\begin{description}
\item[\underline{$\xp =\yp =0$}:] \mbox{\ \ } \\
  Here, the beam axis and the detector symmetry axis are parallel, \mmm\ is the unit matrix and we
  can write :
  \begin{equation}
    \begin{array}{rcl}
      R \sin\phi_i - \yb\ & = & z_i \sin\tilde{\phi_i}\tan\tilde{\theta_i}\\
      R \cos\phi_i - \xb\ & = & z_i \cos\tilde{\phi_i}\tan\tilde{\theta_i}.\\
    \end{array}
    \label{00}
  \end{equation}
  Dividing both equations and using coplanarity
  ($\tan\tilde{\phi_1}=\tan\tilde{\phi_2}$) we can write after some
  algebra exploiting trigonometric relations:
  \begin{equation}
    R\cos\frac{\phi_1-\phi_2}{2} = \xb \cos\frac{\phi_1+\phi_2}{2} +
    \yb \sin\frac{\phi_1+\phi_2}{2} 
    \label{easy}
  \end{equation}
  which is the same as \Eq{dscalesimple} when applying the sign
  convention for $d$ and \ddp\ as explained in Section~\ref{parallel}.

\item[\underline{$\xb = \yb = \yp = 0$, $\xp \neq 0$}:] \mbox{\ \ } \\
  Here we derive:
  \begin{equation}
    \begin{array}{rcl}
      R \cot\phi_i & = & \dsp\sqrt{1-\xps} \cot\tilde{\phi_i} +
      \frac{\xp}{\tan\tilde{\theta_i}\sin\tilde{\phi_i}}\\
      \dsp\frac{z_i}{R \sin\phi_i} & = & \dsp\xp\cot\tilde{\phi_i} + \frac{\sqrt{1-\xps}
        }{\tan\tilde{\theta_i}\sin\tilde{\phi_i}}\\
    \end{array},
    \label{01}
  \end{equation}
  eliminating $\tilde{\theta_i}$ yields
  \begin{equation}
    \sqrt{1-\xps}\cot\phi_i - \frac{\xp\ z_i}{R \sin\phi_i} =
    (1-2\xps) \cot\tilde{\phi_i}.
    \label{01a}
  \end{equation}
  Coplanarity forces $\cot\tilde{\phi_1} = \cot\tilde{\phi_2}$
  and after some more algebra we obtain:
  \begin{equation}
    \begin{array}{rl}
    \dsp R\cos\frac{\phi_1-\phi_2}{2} = & \dsp
    \frac{\xp}{\sqrt{1-\xps}}\left\{\frac{z_1+z_2}{2}
    \cos\frac{\phi_1+\phi_2}{2} - \right.\\&\dsp\left. 
\frac{z_1-z_2}{2}\sin\frac{\phi_1+\phi_2}{2}
    \cot\frac{\phi_1-\phi_2}{2}\right\} 
    \end{array}
    \label{01b}
  \end{equation}
  Keeping only terms linear in \xp\ and considering that
  $\frac{\phi_1-\phi_2}{2} \approx \pm\frac{\pi}{2}$ for elastic
  scattering such that the cotangent is close to zero we may write:
  \begin{equation}
    R\cos\frac{\phi_1-\phi_2}{2} =
    \xp \frac{z_1+z_2}{2} \cos\frac{\phi_1+\phi_2}{2}. 
    \label{01c}
  \end{equation}
    An analog expression is straightforward to derive for $\xp = 0$
  and $\yp \neq 0$.
\end{description}
Taking the definition of \ddb\ and \ddp\ from Section~\ref{parallel} we may
now guess the dependence of \ddb\ on \xb , \yb , \xp , and \yp :
\begin{equation}
  \ddb = \left(\xb + \frac{z_1+z_2}{2} \xp\right) \cos\ddp + \left(\yb + \frac{z_1+z_2}{2} \yp\right) \sin\ddp.
  \label{02}
\end{equation}
This result has been checked numerically to be accurate within 0.2~mm
for reasonable values of beam parameters($|\xb |,|\yb |<10$~mm,$|\xp
|,|\yp |<10$~mrad). 

\subsection{Dependence of \dds\ on Beam Parameters}
The width of the distribution in $d$ at a given angle \ddp\ is derived
simply by using error propagation from \Eq{02}.
\begin{eqnarray}
\dds{}^2 & = & \left(\frac{\partial\ddb}{\partial\xb}\sigma_{\xb}\right)^2 + 
\left(\frac{\partial\ddb}{\partial\yb}\sigma_{\yb}\right)^2 + 
\left(\frac{\partial\ddb}{\partial\xp}\sigma_{\xp}\right)^2 + 
\left(\frac{\partial\ddb}{\partial\yp}\sigma_{\yp}\right)^2 \nonumber\\
& & + \left(\frac{\partial\ddb}{\partial z_1}\sigma_{z_1}\right)^2 +
\left(\frac{\partial\ddb}{\partial z_2}\sigma_{z_2}\right)^2 +
\left(\frac{\partial\ddb}{\partial\ddp}\sigma_{\ddp}\right)^2 
  \label{sigmad}
\end{eqnarray}
We derive
\begin{eqnarray}
  \dds{}^2 & = &
  \left(\sigma^2_{\xb} + \frac{(z_1+z_2)^2}{4} \sigma^2_{\xp}\right) \cos^2\ddp +
  \left(\sigma^2_{\yb} + \frac{(z_1+z_2)^2}{4} \sigma^2_{\yp}\right) \sin^2\ddp
  \nonumber\\
  & & +
  \left(\xps\cos^2\ddp + \yps\sin^2\ddp\right)\frac{\sigma_{z}^2}{2}
    \\
    & &  + \left\{\yb\cos\ddp -\xb\sin\ddp + \frac{z_1+z_2}{2}
    (\yp\cos\ddp - \xp\sin\ddp)\right\}^2 \sigma_{\ddp}^2, \nonumber
    \label{sigmad2} 
  \end{eqnarray}
where the error in $z_1$ and $z_2$ was assumed to be the same. Here,
the error of \ddp\ is given by the detector resolution in the azimuth
\begin{equation}
  \sigma_{\ddp} = \frac{\sigma_\phi}{\sqrt{2}}
  \label{sigmad3}
\end{equation}
Of all these terms most are negligible. Considering that \xb
, and \yb\ are of the order mm, \xp , and \yp\ of the order mrad,
$z_i$ of the order 400~mm, $\sigma_z$ of the order 10~mm, and
$\sigma_{\ddp}\approx$~10~mrad, the last two terms are suppressed by
two orders of magnitude with respect to the leading terms proportional
to $\sigma^2_{\xb}$ and $\sigma^2_{\yb}$. 

The width in \xb\ and \xp\ are related by the amplitude-function
$\beta_x$ of the COSY-ring at the target location
 \begin{equation}
   \sigma_{\xp} = \frac{\sigma_{\xb}}{\beta_x} 
   \label{beta}
 \end{equation}
provided we are at the beam waist.
At the EDDA target location $\beta_x$ and $\beta_y$ is of the order of
 some meters, such that we expect $\sigma_{\xp}$ and
$\sigma_{\yp}$ to be less than a mrad. Therefore the terms proportional
 to $\sigma^2_{\xp}$ and $\sigma^2_{\yp}$ will be a small correction to
 $\sigma^2_{\xb}$ and $\sigma^2_{\yb}$ and we write
\begin{equation}
  \dds{}^2  \approx
  \sigma^2_{\xb} \cos^2\ddp +\sigma^2_{\yb} \sin^2\ddp .
  \label{sigmad4}
\end{equation}
This is only the dependence which enters through the right-hand-side
of \Eq{02}. The quantity \ddb\ itself is computed from
the azimuthal angles which introduces another error through
\Eq{ddef}. This modifies \Eq{sigmad4} to:

\begin{equation}
  \dds{}^2 \approx
  \sigma^2_{\xb} \cos^2\ddp +\sigma^2_{\yb} \sin^2\ddp +
  \frac{R^2\sigma^2_\phi}{2} 
  \label{sigmad5}
\end{equation}
where we have used that $\sin\frac{\phi_1-\phi_2}{2}\approx 1$ for
coplanar events.
This expression, however, is not suitable for a fit. The three
parameters $\sigma_{\xb}$,  $\sigma_{\yb}$, and $\sigma_\phi$ are
highly correlated, i.e. a combination of two can mock up the influence
of the third. This is easy to see, if one multiplies the last term
with $1=\sin^2\ddp + \cos^2\ddp$. The assumption of a perfect angular
resolution ($\sigma_\phi = 0$) 
has the same effect as an increased size of the vertex area both in $x$
and $y$. For a fiber target as it is in use with EDDA the vertex
distribution is very narrow in $y$ (some $\mu$m) so that without loss
of accuracy $\sigma_{\yb}$ may be set to zero and we obtain:
\begin{equation}
  \dds{}^2 = 
  \sigma^2_{\xb} \cos^2\ddp +  \frac{R^2\sigma^2_\phi}{2} 
  \label{sigmad6}
\end{equation}
with two linearly independent parameters $\sigma_{\xb} $ and $\sigma_\phi$.
Alternatively, if the angular resolution is known from other sources, 
$\sigma_{\xb}$ and $\sigma_{\yb}$ may be extracted by this method.


\end{document}